**Barks and Bites at Night: A Study on Free-ranging Dog Behaviour and Human Perception**

Srijaya Nandi[a], Padmanavo Bhuiyan[b], Deepyaman Chakraborty[c], Arun Singh[d], Anindita Bhadra[a*]

[a]Department of Biological Sciences, Indian Institute of Science Education and Research, Kolkata, Mohanpur, Nadia, West Bengal, India

[b]Department of Biochemistry and Biophysics, University of Kalyani, Kalyani, Nadia, West Bengal, India

[c]Indian Association for the Cultivation of Science, Jadavpur, Kolkata, West Bengal, India

[d]Bidhannagar Government College, Salt Lake, Kolkata, West Bengal, India

**Corresponding author:**

Anindita Bhadra

abhadra@gmail.com

**ORCID IDs:**

Srijaya Nandi: https://orcid.org/0000-0002-9462-2374

Padmanavo Bhuiyan: https://orcid.org/0009-0008-8453-1592

Deepayaman Chakraborty: https://orcid.org/0009-0005-1205-6866

Arun Singh: https://orcid.org/0009-0007-1133-7994

Anindita Bhadra: https://orcid.org/0000-0002-3717-9732

**Abstract**

Free-ranging dogs constitute the majority of the global dog population and live in close proximity to humans largely in the Global South. While this coexistence provides some ecological benefits, it also causes dog-human conflict. Anecdotal reports suggest that free-ranging dogs are more aggressive at night, but scientific evidence remains limited. In this study, we examined temporal and social influences on aggression towards humans in free-ranging dogs and compared these patterns with human perceptions and experiences. We conducted censuses across five locations in West Bengal, India, during afternoon (16:00-18:00h) and night (23:00-01:00h), recording 1,246 dogs. 500 structured surveys were conducted with local residents to understand their experiences with and perceptions of aggression from free-ranging dogs. Dogs were more often sighted in groups of three or more at night and more often alone in the day. They were more likely to notice the unfamiliar experimenter and display aggressive behaviour at night, irrespective of whether they were alone or with other dogs, but in general, solitary dogs were less likely to show aggression than dogs in groups. Most respondents who reported a diurnal difference in aggression perceived free-ranging dogs to be more aggressive at night, consistent with the results from the census. However, reported negative interactions were significantly more common during the day, likely reflecting higher human outdoor activity during these hours. Men reported more negative experiences, and individuals who fed dogs were less likely to report such encounters. This study highlights the importance of temporal and social context in shaping aggression in free-ranging dogs and reveals a mismatch between perceived and experienced incidence of dog aggression. This provides important insights into free-ranging dog behaviour that can help to mitigate dog-human conflict.

**Keywords:**

Free-ranging dogs, *Canis lupus familiaris*, dog-human conflict, nocturnal aggression

## Introduction

A large number of animals share spaces with humans in multiple habitats, from highly populated and anthropogenically modified urban to rural areas, and even remote, sparsely inhabited landscapes like forest fringes and protected areas. Such co-habitation often leads to situations of human-animal conflict, which is defined as any interaction that produces a negative consequence for people, animals or both (Schell et al., 2020). Negative consequences for humans include the rapid spread of zoonotic diseases, physical attacks and damage to property and livestock (Bombieri et al., 2018), trauma and stress. Conflict can occur between human and wildlife like tigers, leopards, elephants, bears, due to increasing habitat loss and encroachment of humans on forested areas (Acharya et al., 2017; Ranjan et al., 2025). In such cases, the conflict management is guided by the principles of conservation. However, conflict between humans and several urbanized species like coyotes, bears, cats, dogs, pigeons, and even bees are on the rise, and need inclusive policies based on scientific understanding of the behavioural ecology of the concerned species and their interactions with humans.

In India and many other countries of the Global South, free-ranging dogs (*Canis lupus familiaris*) are commonly found to share myriad human-dominated spaces, from the metropolis to the remotest villages (Biswas et al., 2025; Silva et al., 2022). They form the largest component of the population of domesticated dogs of the globe, and represent an important component of the long-standing ecological and social relationship between humans and dogs. These dogs are typically scavengers, dependent on human-generated waste for their survival and often interact with people actively, developing social bonds with them (Bhattacharjee et al., 2017; Nandi et al., 2026). Dogs are a rare urban-adapted species that prefers to give birth in the vicinity of human activity, rather than avoiding anthropogenic disturbance completely (Sen Majumder et al., 2016).

While this close association with humans provides ecological advantages for dogs, it also leads to several challenges. Free-ranging dogs are often perceived as a nuisance because they scatter garbage, defecate in public spaces, and engage in nocturnal barking (Bhalla et al., 2021; Bhattacharjee et al., 2017; Ratsitorahina et al., 2009). More importantly, they serve as reservoirs of zoonotic diseases such as rabies and can transmit the infection through bites, posing a significant public health concern (Beron et al., 2026). One of the unintended byproducts of urban expansion is the sharp increase in free-ranging dog populations around human settlements, largely driven by greater availability of food sources such as unmanaged waste and direct human feeding (Büyük et al., 2025; Hughes & Macdonald, 2013). The global free-ranging dog population is estimated to be around 200 million (Chandran & Azeez, 2016), of which India is projected to host an around 59 million (Gompper, 2014). However, these estimates are based on projections of local surveys and vaccination clinic data, and need revisiting. Nevertheless, dog-human conflicts have been increasingly reported from Indian and other countries in the recent years (Kumar, 2026). A recent estimate suggested roughly 12 million stray dog bite incidents per year in India (Bhalla et al., 2021). Despite the magnitude of this issue, many measures implemented by municipal authorities remain largely reactive and often lack scientific grounding. While population control measures and vaccination programmes are important components of management strategies, there is an urgent need for systematic scientific research examining the dynamics of dog-human conflict. In 2025, the Supreme Court of India issued certain guidelines for the management of stray dog populations across the country, which have faced much resistance from dog welfare organizations, caregivers and dog lovers. The guidelines were issued following the court's Suo moto cognizance of newspaper reports of dog bites, and this case created much distress for citizens who are concerned about the wellbeing of the animals and their rights, leading to a sharply divided public opinion (Prasad, 2026). This case alone suggests the urgent need for policies based on a deeper understanding of free-ranging dog behaviour and human interaction patterns through rigorous scientific enquiry.

Most studies conducted on wild mammals report increased nocturnality as a result of human disturbance, in order to avoid direct human contact (Gaynor et al., 2018; Procko et al., 2023). However, instances of aggression in them have not been reported to be higher at night compared to daytime. In contrast, the activity patterns of Indian free-ranging dogs align well with those of humans living alongside them (Banerjee & Bhadra, 2021). It is not however known whether they exhibit differential tendencies of aggression towards humans during the diurnal and nocturnal periods of the day. Some news articles have suggested that these dogs exhibit heightened aggression during night hours (*The New Indian Express*, 2025), but this claim has not yet been scientifically validated at the population level.

Further, Indian free-ranging dogs tend to be bolder in the presence of their group members as opposed to when present alone. This is evident in higher rates of approach and lesser avoidance behaviour displayed when presented with threatening cues (Bhattacharjee et al., 2020). The time of day has also been found to determine the association in dogs, with free-ranging dogs in India scavenging alone (Sen Majumder et al., 2014) and Andean dogs begging alone (Finzi et al., 2023) during daytime. It is not known whether the type of association or the type of day or both affect the levels of aggression towards humans in free-ranging dogs.

Human-animal conflict is sometimes influenced by fear and perception of people towards the concerned species. For example, snakes are considered a threat by most humans, irrespective of whether they are venomous or not (Babalola et al., 2020). In India, dogs on the streets often elicit widely different reactions from people – from the overtly loving and caring who feed, pet and care for dogs, to the tolerant but not overtly friendly, to the aversive, who either are fearful of dogs, or have a deep aversion towards them, leading to agnostic behaviours like threatening, beating and even killing of dogs. Hence, while carrying out a study on dog-human conflict, in addition to understanding dog behaviour towards humans in different contexts, we need to also understand human perceptions of the free-ranging dogs in the same habitats. We conducted a study to examine how aggression in free-ranging dogs might vary between diurnal and nocturnal hours of the day, and with the number of dogs present together. Additionally, we carried out surveys among local residents to understand their experiences with and perceptions of aggression displayed by free-ranging dogs. We hypothesised that (i) the number of dogs sighted will not differ between the day and night, while the type of association will, with fewer solitary dogs being sighted during the night; (ii) free-ranging dogs would exhibit higher levels of aggression during nighttime, and that aggression would be more common when dogs are in a group; and (iii) people will consider dogs to be more aggressive during nighttime.

## Methods

### Ethics Statement

The study design did not violate the Animal Ethics regulations of the Government of India (Prevention of Cruelty to Animals Act 1960, Amendment 1982). The experimental protocol was approved by the IISER Kolkata Animal Ethics Committee, as part of a larger project sanctioned by the SERB (EMR/2016/000595).

## Safety Measures

Prior to the start of fieldwork, the experimenters involved in the spot census were administered pre-exposure prophylactic rabies vaccination in accordance with public health recommendations for individuals working closely with free-ranging dogs.

## Study Sites

The spot census and the structured surveys were conducted across five distinct locations within the Nadia district of West Bengal, India: Gayeshpur (22°57'29"N, 88°29'33"E), Kalyani (22°58'29"N, 88°26'37"E), Kataganj (22°56'58"N, 88°28'14"E), Jaguli (22°55'38"N, 88°32'46"E) and Anandanagar (22°58'44"N, 88°29'15"E). These locations are represented in Figure S1. Four non-overlapping zones were selected within each location for the study.

## Experimental Procedure

The observations were conducted by two male experimenters, both unfamiliar with the local dog populations. The experimenters walked along roads in the designated zone for two hours, both during afternoon (16:00-18:00 hours) and night (23:00-01:00 hours). Experimenter 1 (E1) walked through the designated zone and attempted to establish eye contact with any free-ranging dog encountered. Eye contact was considered successful only when the dog returned the gaze. Experimenter 2 (E2) followed E1 at an approximate distance of 2 meters and video recorded the dogs' behaviours upon noticing E1, as well as their initial states (prior to noticing E1). If any dog (whether solitary or part of a pair or group) displayed aggressive behaviour, E1 responded by revealing a stick that was otherwise kept concealed in his trouser pocket. All the dogs in a group or both in the pair were exposed to the stick even if they themselves did not behave aggressively, if at least one dog in the unit exhibited aggression. In cases where dogs displayed aggression a few seconds after E1 had walked past them, the stick was not revealed. Dogs sighted within a maximum distance of 5 meters of each other and not displaying aggression towards one another were considered part of the same unit.

Each zone was visited twice: once during the afternoon (16:00-18:00 hours) and once at night (23:00-01:00 hours). The census sessions were carried out between May and July 2025. The order of time slots in which each location was visited was randomly assigned. A minimum interval of five days was maintained between visits to the same location to reduce familiarity bias. The roads covered on foot were recorded using the Strava app on a mobile device. This enabled us to cover the same roads during the second visit to a given location. During the nighttime walk, the experimenters covered a torch (Philips Torch, 1500 Lumen) with three layers of red cellophane paper and used it to illuminate the road, providing sufficient visibility while preserving low-light conditions. Figure S2 shows the route that was followed by the experimenters on foot during the two-hour period of the spot census on one of the observation days.

**Surveys**

A total of 500 structured, in-person surveys were conducted at the same locations where the spot census had previously been carried out. Table S1 contains the details of the number of surveys that were conducted in each area within each location. The surveys targeted households, with one willing adult respondent selected from each household. Households were randomly selected within each zone, and informed consent was obtained from all participants prior to data collection. A participant must have been a permanent resident of the locality for at least the last five years in order to be included. The surveys were conducted between November 2025 and February 2026, between 11:00 and 17:00 hours.

The questionnaire included in the survey is listed in Table S2.

**Data coding and analysis**

Each video included annotations regarding the dog's sex and approximate age. However, for some of the dogs, the sex could not be visually determined, especially for the ones that ran away on seeing the experimenters. For such dogs, the sex was coded as “ui”, meaning unidentified. Dogs were classified as adults, juveniles or pups, based on their body size and the appearance of nipples in females and testes in males, with adults having descended testes and darker nipples. Pups were typically 0-3 months of age while juveniles were 3-9 months. Since the age classification was done based on the visual characteristics, some misclassification might have been possible. The demeanours of the dogs were classified into four types: neutral, affiliative, anxious and aggressive. Affiliative demeanour included tail wag, anxious demeanour included tail droop, crouched posture, aggressive demeanour included barking, growling and chasing and finally the neutral demeanour included all behaviours apart from the ones mentioned in the three other demeanour categories. A second coder coded 20% of randomly selected videos and the inter-observer reliability for the demeanours was found to be high ($\kappa = 0.88$).

All statistical analyses were conducted using R version 4.2.3 (R Core Team, 2023) in RStudio. A Chi-squared goodness-of-fit test was performed to compare dog abundance between the two sampling times. A multinomial logistic regression (Venables & Ripley, 2002) was used to test whether time of day predicted group size type (solitary: a dog sighted alone; pair: 2 dogs sighted together; group: 3 or more dogs sighted together). In total, 719 dog units comprising 1,246 dogs were recorded. Group size type was treated as a categorical response variable, and time of day (afternoon/night) was included as a predictor. Binomial logistic regressions (glm in base R for fixed-effects models and lme4 for mixed-effects models; Bates et al., 2015) were fitted to examine whether time of day predicted the likelihood of a dog noticing the unfamiliar experimenter, as well as whether the dog unit or individual dogs displayed aggression. In the model examining aggression at the dog-unit level, time of day and group size type were included as predictors. In the second model, which tested aggression at the individual level, age and sex were included as additional predictors along with time of day and group size type. A multinomial logistic regression was conducted to examine whether dogs' demeanour when seeing the experimenter (affiliative, aggressive, anxious) varied as a function of time of day (afternoon vs. night) and group size type (group, pair, solitary), with neutral demeanour as the reference category. A Chi-squared goodness-of-fit test was performed to compare the number of dogs that did and did not show aggression after being exposed to the stick, among dogs that were not aggressive on seeing the experimenter but were shown the stick because another dog in their unit was aggressive.

The results from the survey questions were converted into proportions, and Chi-squared goodness-of-fit tests were performed whenever necessary. Age class and education levels were clubbed together to obtain a substantial number of respondents in each category. A binomial logistic regression was conducted to examine predictors of reporting negative experiences with free-ranging dogs. The model included gender, age class (18-39, 40-59 or 60+ years), education level (up to Class 10, up to Class 12, or graduation and above), pet ownership, feeding behaviour, petting behaviour, and care practices towards free-ranging dogs as predictors. Age class and education level were treated as categorical predictors.

Mixed-effects and fixed-effects models were compared using likelihood ratio tests; the fixed-effects model was retained when the random effect of location did not improve model fit. For the dog-unit aggression model, an interaction between time of day and group size type was also tested with a likelihood ratio test, to check whether the effect of time of day differed between group size types; it did not improve the model and was not retained. Model diagnostics, including checks for model fit, multicollinearity, dispersion, and residual structure, were carried out using the performance (Lüdecke et al., 2021) and DHARMa (Hartig, 2024) packages. Post-hoc comparisons and estimated marginal means were obtained using emmeans (Lenth & Piaskowski, 2026). Post hoc pairwise comparisons (Tukey-adjusted) were carried out only for categorical predictors that had three or more levels and at least one significant coefficient in the main model; this applied to group size type (dog-unit aggression, individual aggression and demeanour models) and the sex of the dogs. Predictors with two levels were interpreted directly from the model coefficients, and no pairwise comparisons were made for the age of the dogs or the age class and education level of the survey respondents, as none of their coefficients were significant. For the multinomial model of demeanour, we compared the predicted probability of each demeanour (neutral, affiliative, aggressive and anxious) between each pair of group size types, using Tukey-adjusted asymptotic z-tests. Visualisations were generated using ggplot2 (Wickham, 2016) and ggthemes (Arnold, 2025). An alpha level of 0.05 was used for all the analyses conducted.

## Results

### A) Spot census

#### 1) Effect of time of day on dog abundance

In total, 1,246 dogs were recorded (afternoon: 614; night: 632), and dog abundance did not differ significantly between the afternoon and nighttime hours ($\chi^2 = 0.260$, df = 1, $p = 0.610$).

#### 2) Effect of time of day on group size type

Of the 719 dog units recorded, 404 were solitary dogs, 182 were pairs and 133 were groups. Compared to groups, the odds of sighting pairs ($b$ = -0.50, SE = 0.23, $z$ = -2.16, $p$ = 0.031) and solitary dogs ($b$ = -0.76, SE = 0.20, $z$ = -3.74, $p$ < 0.001) were lower at night than in the afternoon. Groups were more common at night (24.0%) than in the afternoon (13.8%), whereas solitary dogs were less common at night (50.0%) than in the afternoon (61.6%; Figure 1).

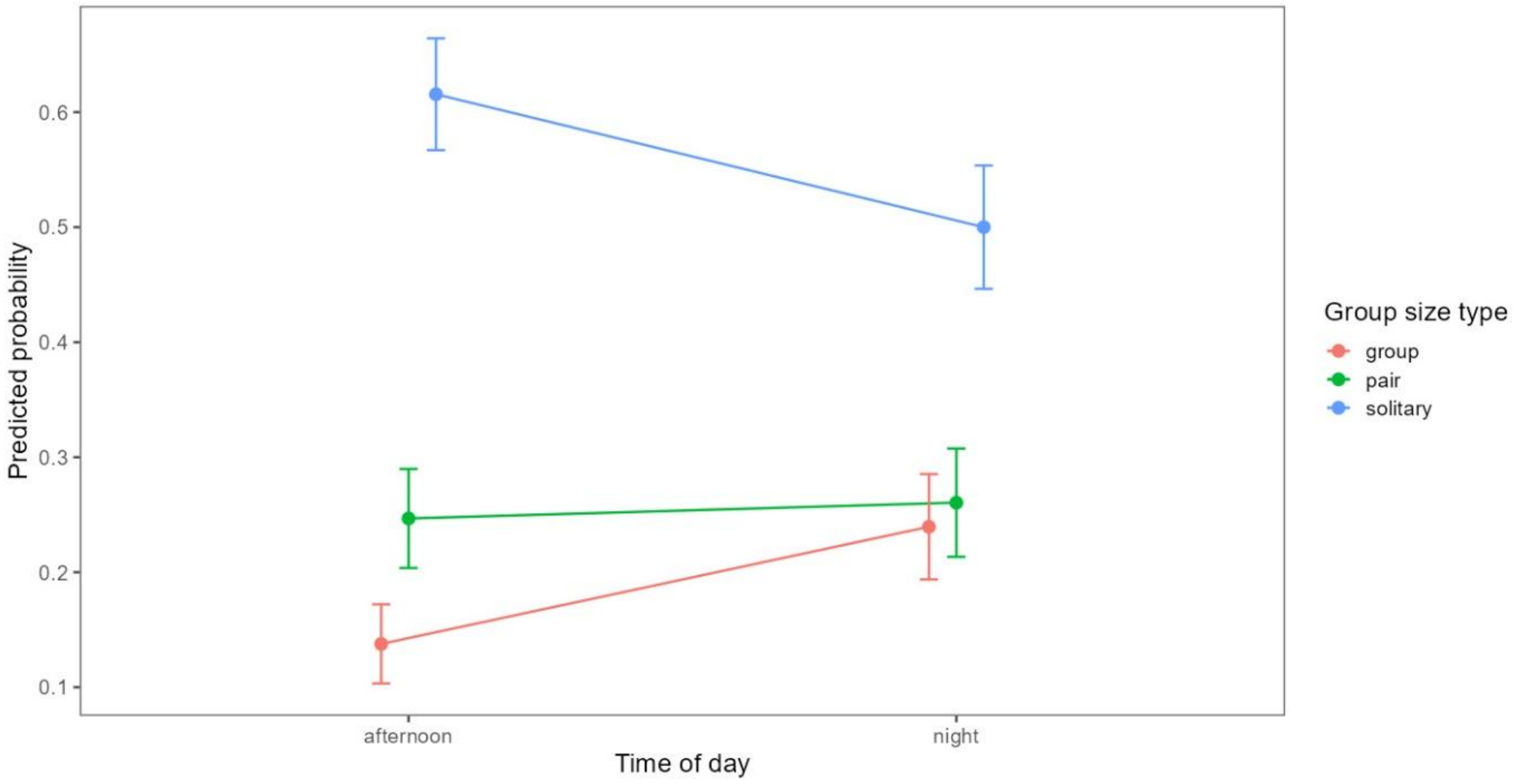

**Figure 1.** Predicted probabilities of different group size types observed during the afternoon (16:00-18:00 h) and night (23:00-01:00 h). Error bars show 95% confidence intervals.

**3) Effect of time of day on whether dogs noticed the unfamiliar experimenter**

Including location as a random effect did not improve this model ($\chi^2$ = 2.00, df = 1, $p$ = 0.158), so the fixed-effects model is reported. The effect of time of day was significant, such that the dogs had a higher likelihood of noticing the unfamiliar experimenter at night than in the afternoon (73.3% vs 50.2% of dogs; $b$ = 1.001, SE = 0.121, $z$ = 8.289, $p$ < 0.001).

**4) Effect of time of day and number of dogs present together on the dog unit's tendency to show aggression**

Including location as a random effect did not improve the dog-unit model ($\chi^2$ = 0.79, df = 1, $p$ = 0.374), so the fixed-effects model is reported. Aggression was more likely during the night than the afternoon ($b$ = 1.409, SE = 0.230, $z$ = 6.136, $p$ < 0.001), with about four times higher odds of aggression at night after accounting for group size type (OR = 4.09, 95% CI 2.61-6.42). This effect did not differ between group size types (time of day × group size type interaction: $\chi^2$ = 1.36, df = 2, $p$ = 0.506); aggression was higher at night than in the afternoon in solitary dogs (19.2% vs 5.5% of units), pairs (36.8% vs 9.5%) and groups (32.5% vs 15.1%). Group size type also significantly predicted aggression. Compared to dogs in groups, solitary dogs were significantly less likely to exhibit aggression ($b$ = -0.813, SE = 0.263, $z$ = -3.094, $p$ = 0.002). However, dogs in pairs did not differ significantly in their aggression levels compared to dogs in groups ($b$ = -0.010, SE = 0.278, $z$ = -0.036, $p$ = 0.971). Post hoc pairwise comparisons (Tukey-adjusted) revealed a significant effect of group size on aggression. Solitary dogs had significantly lower odds of aggression compared to both groups (OR = 2.25, SE = 0.593, $z$ = 3.094, $p$ = 0.006) and pairs (OR = 2.23, SE = 0.548, $z$ = 3.270, $p$ = 0.003). There was no significant difference in aggression between groups and pairs (OR = 1.01, SE = 0.280, $z$ = 0.036, $p$ = 0.999). Figure 2 represents the predicted probability of aggression at the level of the dog unit.

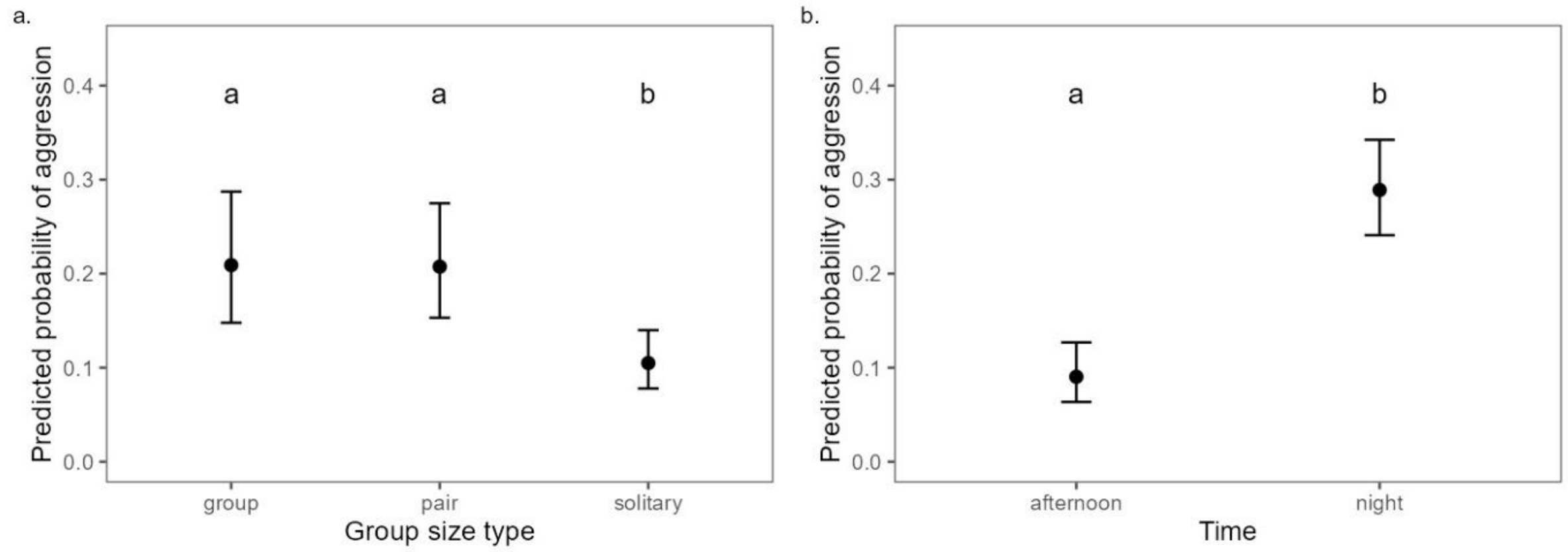

**Figure 2.** Predicted probability of aggression at the level of the dog unit across (a) different group size types and (b) different times of the day. Error bars show 95% confidence intervals; levels sharing a letter do not differ significantly (Tukey-adjusted comparisons for group size type; model coefficient for time of day).

**5) Effect of time of day, number of dogs present together, age and sex of dogs on whether the individual dogs showed aggression**

To examine whether group-level patterns were driven by individual behavioural tendencies, we conducted an additional analysis at the individual level.

There was a significant effect of time of day, such that aggression was more likely during the night (*b* = 1.32, SE = 0.18, *z* = 7.48, $p < 0.001$). Sex was a significant predictor only because of the unidentified ("ui") category. Individuals in the "ui" category were more likely to exhibit aggression compared to the female dogs (*b* = 0.93, SE = 0.38, *z* = 2.49, *p* = 0.013), whereas males did not differ significantly from females (*b* = -0.145, SE = 0.167, *z* = -0.871, *p* = 0.384). Because sex could not be identified mainly for dogs that ran away on seeing the experimenters, this effect is unlikely to reflect a true difference between the sexes, therefore post-hoc comparison was not performed. Group size type showed a significant effect, with solitary individuals being less likely to display aggression than dogs in groups (*b* = -0.677, SE = 0.203, *z* = -3.334, $p < 0.001$), while pairs did not differ significantly (*b* = -0.176, SE = 0.188, *z* = -0.936, *p* = 0.349). Age was not a significant predictor of aggression (juveniles vs adults: *b* = -0.160, SE = 0.190, *z* = -0.844, *p* = 0.399; pups vs adults: *b* = -0.289, SE = 0.574, *z* = -0.504, *p* = 0.614). The random effect of location indicated modest variability in baseline aggression across sites (SD = 0.31), and including it improved model fit ($\chi^2 = 6.07$, df = 1, *p* = 0.014).

Post-hoc pairwise comparisons examining differences in aggression across group size types revealed that solitary dogs had significantly lower odds of exhibiting aggression compared to groups (OR = 1.968, SE = 0.400, *z* = 3.334, *p* = 0.003). There was no significant difference in aggression between groups and pairs (OR = 1.192, SE = 0.224, *z* = 0.936, *p* = 0.618), nor between pairs and solitary dogs after adjustment for multiple comparisons (OR = 1.650, SE = 0.357, *z* = 2.318, *p* = 0.053). Figure 3 represents the predicted probability of aggression at the individual dog level.

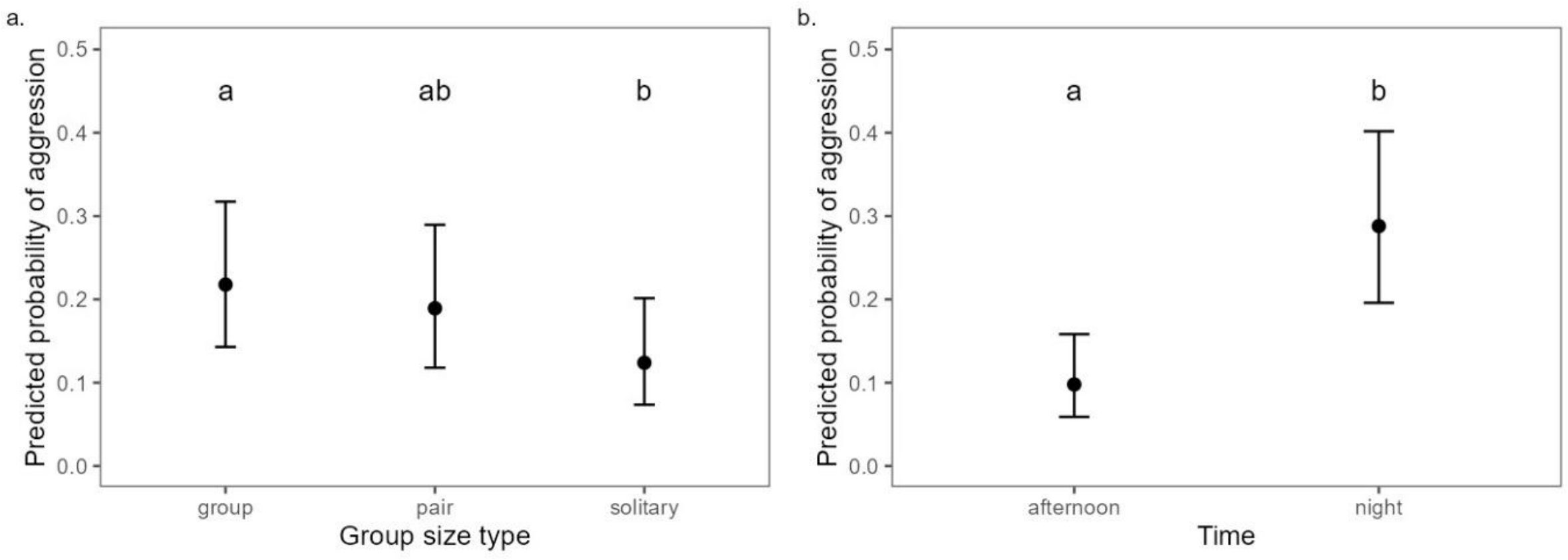

**Figure 3.** Predicted probability of aggression across (a) different group size types and (b) different times of the day at the individual level. Error bars show 95% confidence intervals; levels sharing a letter do not differ significantly (Tukey-adjusted comparisons for group size type; model coefficient for time of day).

**6) Demeanour of the dogs on seeing an unfamiliar person**

Among the 771 dogs that noticed the experimenter, affiliative demeanour was not significantly affected by the time of day ($p = 0.233$) or group size (pair: $p = 0.823$; solitary: $p = 0.364$) relative to neutral. In contrast, aggressive demeanour was significantly more likely at night compared to the afternoon ($b = 0.99$, SE = 0.18, $z = 5.37$, $p < 0.001$) and was less likely in solitary dogs than in groups ($b = -0.83$, SE = 0.21, $z = -3.91$, $p < 0.001$), whereas pairs did not differ significantly from groups ($b = -0.33$, SE = 0.20, $z = -1.69$, $p = 0.091$). Anxious demeanour was also more likely at night ($b = 1.16$, SE = 0.52, $z = 2.21$, $p = 0.027$), but there was no significant effect of group size on anxious demeanour (pair: $p = 0.527$; solitary: $p = 0.165$). Post hoc pairwise comparisons (Tukey-adjusted) of predicted probabilities showed that aggressive demeanour was more likely in groups than in solitary dogs (difference = 0.157, SE = 0.036, $z = 4.34$, $p < 0.001$) and in pairs than in solitary dogs (difference = 0.087, SE = 0.037, $z = 2.37$, $p = 0.047$), but did not differ between groups and pairs ($p = 0.167$). Neutral demeanour was more likely in solitary dogs than in groups (difference = 0.117, SE = 0.039, $z = 3.04$, $p = 0.007$). Affiliative and anxious demeanours did not differ between any group size types (all $p > 0.17$).

**7) The response of dogs to the stick**

All 197 dogs that initially displayed aggression and were then shown the stick continued to behave aggressively. Of these, only one dog escalated the interaction by chasing E1.

Of the 38 dogs that were initially non-aggressive upon seeing the experimenter but were exposed to the stick because another dog present with them behaved aggressively, 2 dogs subsequently displayed aggressive demeanour, 4 dogs exhibited an anxious demeanour, and the remaining 32 dogs retained a neutral demeanour after exposure; Thus, a significantly higher proportion of dogs did not show aggression ($\chi^2 = 30.421$, df = 1, $p < 0.001$).

**B) Surveys**

**1) Do people think free-ranging dogs are more aggressive at certain times of the day?**

A majority of respondents (65%) reported a diurnal difference in aggression levels, while 35% did not. Overall, respondents indicated that free-ranging dogs appeared more aggressive at certain times of day than others, a difference that was statistically significant ($\chi^2 = 45$, df = 1, $p < 0.001$).

**2) At what time of the day do people think that free-ranging dogs are more aggressive?**

The period from 06:00 to 18:00 hours was defined as daytime, while 18:00 to 06:00 hours was defined as nighttime. Respondents perceived free-ranging dogs to be more aggressive during the night than during daytime (431 of 462 time-slot responses; $\chi^2 = 346.32$, df = 1, $p < 0.001$; Figure 4a). Table 1 presents the time slots and the number of responses corresponding to each.

**Table 1.** The frequency of time slots during which respondents perceived dogs to be more aggressive.

| **Time** | **Time slot** | **Total count** |
|---|---|---|
| Nighttime | 18:00-21:00 | 50 |
| | 21:00-00:00 | 203 |
| | 00:00-03:00 | 174 |
| | 03:00-06:00 | 4 |
| Daytime | 06:00-09:00 | 7 |
| | 09:00-12:00 | 7 |
| | 12:00-15:00 | 9 |
| | 15:00-18:00 | 8 |

**3) How likely are people to encounter negative interactions with free-ranging dogs?**

The people surveyed were significantly less likely to report having experienced negative interactions with free-ranging dogs (negative interaction: 28%, no negative interaction: 72%; $\chi^2 = 96.800$, df = 1, $p < 0.001$).

**4) During the experienced negative interaction, were the free-ranging dogs more likely to be solitary or in the presence of other dogs?**

Among the respondents who reported negative interactions with free-ranging dogs, 50.71% of the cases concerned free-ranging dogs that were solitary, while in the remaining 49.29% of the cases, the involved dog was in the presence of other dogs, indicating no significant difference in the probability that these dogs were or were not alone ($\chi^2 = 0.028$, df = 1, $p = 0.866$).

**5) At what time of the day were the experienced negative interactions more common?**

Out of the 140 respondents who reported to have experienced negative interactions with free-ranging dogs, 16 were not sure about the timings of the events. Among the remaining respondents, the encountered negative interactions were significantly more common during daytime compared to nighttime (88 of 143 time-slot responses; $\chi^2 = 7.615$, df = 1, $p = 0.006$; Figure 4b). Table 2 presents the time slots and the number of responses corresponding to each.

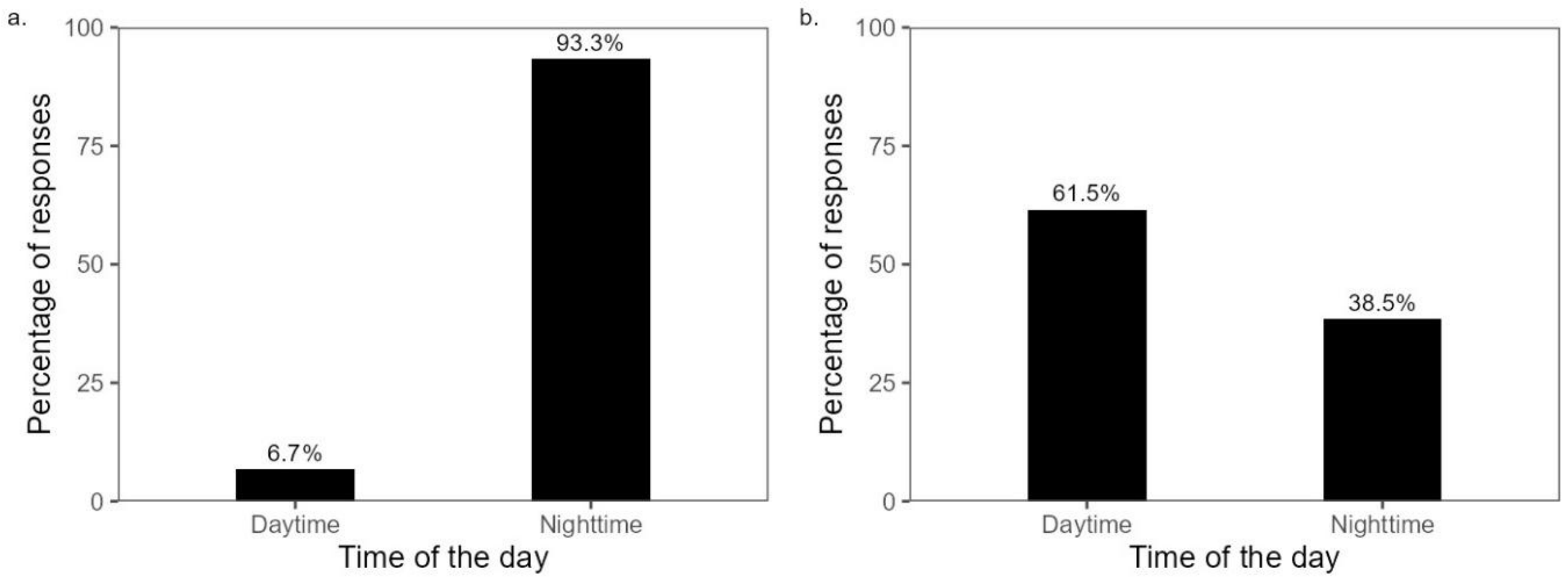

**Figure 4.** Influence of time of day on (a) perceived aggression and (b) experienced negative interactions with free-ranging dogs.

**Table 2.** The frequency of time slots during which the respondents encountered more negative interactions with free-ranging dogs.

| Time | Time slot | Total count |
|---|---|---|
| Nighttime | 18:00-21:00 | 28 |
| | 21:00-00:00 | 17 |
| | 00:00-03:00 | 9 |
| | 03:00-06:00 | 1 |
| Daytime | 06:00-09:00 | 10 |
| | 09:00-12:00 | 30 |
| | 12:00-15:00 | 26 |
| | 15:00-18:00 | 22 |

**6) What kind of negative interaction did people experience?**

Of the 140 respondents who reported experiencing negative interactions with free-ranging dogs, 70 (50%) reported being bitten, while the remaining 70 (50%) described other forms of negative interactions, including scratching, chasing, barking, growling, or attempted biting.

**7) Factors affecting the chances of experiencing a negative interaction with a free-ranging dog**

Male respondents were significantly more likely to report negative experiences than females ($b = 0.83$, SE = 0.22, $z = 3.72$, $p < 0.001$; Figure 5a). In contrast, people feeding dogs were associated with a lower likelihood of negative experiences ($b = -0.80$, SE = 0.32, $z = -2.50$, $p = 0.012$; Figure 5b). Other predictors, including pet ownership and whether the person petted or cared for free-ranging dogs, did not significantly influence the probability of experiencing negative interactions (all $p > 0.05$). Neither age class (40-59 vs 18-39 years: $b = 0.01$, SE = 0.24, $z = 0.05$, $p = 0.959$; 60+ vs 18-39 years: $b = -0.16$, SE = 0.31, $z = -0.51$, $p = 0.609$) nor education level (up to Class 12 vs up to Class 10: $b = -0.26$, SE = 0.29, $z = -0.91$, $p = 0.364$; graduation and above vs up to Class 10: $b = -0.23$, SE = 0.25, $z = -0.90$, $p = 0.370$) significantly influenced negative experiences, so no pairwise comparisons were made between their levels. Including location as a random effect improved this model ($\chi^2 = 4.21$, df = 1, $p = 0.040$; SD = 0.33), so the mixed-effects model is reported. Respondents who did not report their age (n = 16) were excluded from this model, leaving 484 respondents.

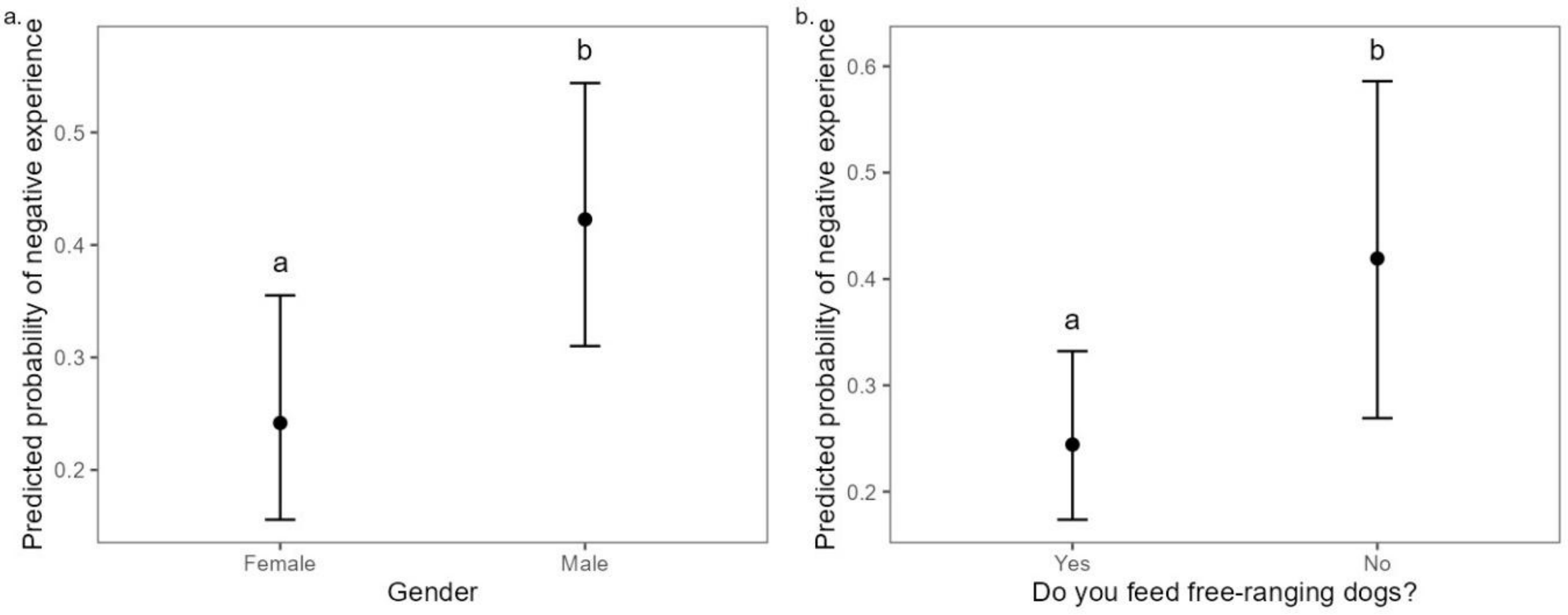

**Figure 5.** Predicted probability of negative experiences with free-ranging dogs based on (a) gender of the respondent and (b) whether the respondent feeds dogs. Error bars show 95% confidence intervals.

## Discussion

The present study examined temporal and social influences on aggression towards humans in free-ranging dogs and highlighted how these behavioural patterns align with local human perceptions and experiences.

The total number of dogs sighted did not differ between the day and night census, in line with our hypothesis. However, the composition of dog units differed between the two census periods. Solitary dogs were most commonly sighted at both times, but groups of three or more dogs were sighted more at night than in the afternoon, while solitary dogs were sighted less at night. This pattern is consistent with studies of Andean dogs, which report solitary begging during the day and coalition formation at night (Finzi et al., 2023), and Sen Majumder et al. (2014), who found that dogs scavenge mostly alone during the daytime. Increased nocturnal vigilance and territoriality may explain why dogs showed a higher probability of noticing the unfamiliar experimenter at night.

The likelihood of aggression displayed towards the experimenter was significantly influenced by both time of day and the size of the dog unit. Aggression was more frequent at night and was lower in solitary dogs than in pairs or groups. To our knowledge, this represents one of the first pieces of evidence demonstrating diurnal variation in aggression levels directed towards humans, among free-ranging dogs. Although dogs were more often in groups at night, group size type was included in the same model as time of day, so the nocturnal increase in aggression was not simply a consequence of dogs being in larger groups at night. Indeed, aggression was higher at night in solitary dogs, pairs and groups alike, and the size of this increase did not differ between them. The higher probability of aggression when other dogs were present supports the "bolder together" phenomenon previously reported by Bhattacharjee et al. (2020), suggesting that presence of other members of the group may facilitate risk-taking behaviours.

Patterns at the individual level paralleled those observed at the level of the dog unit, although the difference between pairs and solitary dogs was not significant at the individual level. Individual dogs were significantly more likely to display aggression at night and when in groups compared to when alone. Age did not influence aggressive behaviour, and the only effect of sex involved dogs whose sex could not be identified (which included the dogs that ran away on seeing the experimenter), indicating that temporal and social factors play a more prominent role than demographic characteristics of the individual dogs in shaping aggressive responses in this context.

Affiliative demeanour was not affected by time of day or group size. In contrast, aggressive and anxious demeanours were more prevalent at night, and aggressive demeanour was less likely in solitary dogs than in groups or pairs. Together, these findings strengthen the idea that at night, the presence of other dogs may heighten arousal or defensive behaviours in free-ranging dogs, leading to increased aggressive and anxious behavioural responses.

All 197 aggressive dogs that were shown the stick continued to behave aggressively. However, the dogs that were exposed to the stick, not because they themselves showed aggression, but because the other dogs present with them did, showed significantly fewer events of aggression. This suggests that the demeanour displayed by a dog may be influenced more strongly by intrinsic temperamental traits or personalities, than by the immediate external stimulus presented to it.

Our survey responses helped us to explore how people perceive and experience interactions with free-ranging dogs. When respondents were asked whether free-ranging dogs are more aggressive at certain times of the day, a significant majority (65%) reported a time-of-day difference in aggression, and most of these respondents perceived dogs to be more aggressive at night than during the daytime. This perception aligns with our census findings, which demonstrated higher levels of observed aggression during the night. However, a significantly smaller proportion (only 28%) of our surveyed respondents reported having personally experienced negative interactions with free-ranging dogs, but around half of them reported being bitten. Moreover, among those who had experienced negative interactions, such incidents were reported to occur more frequently during the daytime rather than at night, contrary to the behavioural patterns documented in our census. This apparent discrepancy between perception about dog aggression and personal experience is worth noticing. Although aggression was found to be higher at night, people reported more negative encounters during the day. One likely explanation is the differential human activity patterns: people are substantially more active outdoors during the day, increasing the probability of negative human-dog interactions and, consequently, negative encounters. In contrast, nighttime human activity on the streets is relatively limited, reducing opportunities for interaction despite elevated canine aggression levels. Additionally, it is also possible that dogs may encounter fewer humans at night and thus respond more defensively toward unfamiliar individuals during these periods. Therefore, perceived nocturnal aggression may stem from the salience of nighttime encounters, heightened vigilance, or cultural narratives surrounding nighttime risk rather than direct personal experience. It is also worth noting that people did not mention dogs to be more often in groups when experiencing negative interactions with them, which contradicts the findings of this study. Most of the negative encounters between humans and these dogs were reported during daytime and might have been the reason behind this observed discrepancy, since dogs do not tend to be more often in groups during the daytime.

Another notable finding was that male respondents were significantly more likely to report negative interactions with free-ranging dogs compared to the female respondents. This is in line with the findings of Sudarshan et al. (2006), who report men to be bitten by dogs more often compared to women. One possible explanation is differential exposure: in the present surveyed population, men may spend more time outdoors, thereby increasing the likelihood of encountering dogs and experiencing negative interactions. Thus, the observed gender difference may reflect variation in opportunity for interaction rather than differences in dog-directed behaviour. Feeding behaviour also emerged as an important factor. Respondents who reported feeding free-ranging dogs were significantly less likely to report negative interactions than those who did not feed them. Interestingly, other forms of interaction, such as petting dogs or providing care when they were sick, did not significantly influence the likelihood of negative experiences. This pattern suggests that regular feeding may play a significant role in shaping dog-human relationships. Dogs may be more familiar with people who feed them regularly and therefore less likely to respond aggressively towards them. Alternatively, people who choose to feed dogs may be less fearful, more confident in approaching them, or better able to interpret their behaviours, thereby lowering the probability of conflict.

Our study involved the experimenters gazing at the dogs. A further step would be to see if avoiding eye contact with the dogs can reduce the levels of aggression observed. Future research could extend these findings by examining how free-ranging dogs respond to familiar versus unfamiliar humans at night. Such work would help disentangle whether increased nocturnal aggression reflects generalised arousal, territorial defence, reduced habituation to human presence, or context-specific responses to unfamiliar individuals.

Dog-human conflict has been on the rise in recent years, particularly in India, where increasing societal polarisation is evident in perceptions of free-ranging dogs. In this context, a study like ours is of considerable importance, as it provides empirical insight into patterns of dog-directed aggression as well as the perceptions and experiences that the general public have with free-ranging dogs. Our findings also have a practical implication: displaying a stick did not stop dogs that were already aggressive from behaving aggressively, although only one dog escalated to chasing and no bites occurred. Because there was no comparison without the stick, whether showing a stick reduces the risk of bites remains to be tested. A limitation of this study was that we could not emulate more context of conflict like someone attacking a dog or threatening it aggressively, as these acts would be unethical, though dogs do encounter such interactions on the streets of India, and also pose a threat to the experimenter. As such, this study may serve as a foundation for future research aimed at better understanding dog-human conflict, identifying context-specific risk factors associated with dog aggression, and developing evidence-based strategies for conflict mitigation and improved coexistence between humans and free-ranging dogs.

**Acknowledgements**

The authors are thankful to Mr. Imran Mondal for his assistance during fieldwork.

**CRediT authorship contribution statement**

Srijaya Nandi: Conceptualization, Methodology, Formal analysis, Investigation, Data curation, Writing – Original Draft, Visualization, Project administration. Padmanavo Bhuiyan: Investigation, Data curation. Deepyaman Chakraborty: Investigation, Data curation. Arun Singh: Investigation. Anindita

Bhadra: Conceptualization, Methodology, Resources, Writing – Review & Editing, Supervision, Funding acquisition.

**Funding Sources**

This research was supported by the Department of Science and Technology, Ministry of Science and Technology (INSPIRE fellowship to S.N.) and the Janaki Ammal – National Women Bioscientist Award (BT/HRD/NBA-NWB/39/2020-21 (YC-1)), Department of Biotechnology, India.

**Conflict of Interest**

The authors declare that the research was conducted in the absence of any commercial or financial relationships that could be construed as a potential conflict of interest.

**References**

Acharya, K., Paudel, P., Neupane, P. R., & Köhl, M. (2016). Human–wildlife conflicts in Nepal: Patterns of human fatalities and injuries caused by large mammals. *PLOS ONE, 11*. https://doi.org/10.1371/journal.pone.0161717

Acharya, K. P., Paudel, P. K., Jnawali, S. R., Neupane, P. R., & Köhl, M. (2017). Can forest fragmentation and configuration work as indicators of human–wildlife conflict? Evidences from human death and injury by wildlife attacks in Nepal. *Ecological Indicators, 80*, 74-83. https://doi.org/10.1016/j.ecolind.2017.04.037

Arnold, J. (2025). *ggthemes: Extra themes, scales and geoms for 'ggplot2'* (R package version 5.2.0.9000) [Computer software]. https://jrnold.github.io/ggthemes/

Babalola, O., Jegede, H. O., & Ogunro, B. N. (2020). Perceptions, attitudes, and outcomes of human-snake encounters: A retrospective study of an online discussion community in Nigeria. *Asian Journal of Ethnobiology, 3*. https://doi.org/10.13057/asianjethnobiol/y030101

Banerjee, A., & Bhadra, A. (2021). Time-activity budget of urban-adapted free-ranging dogs. *Acta Ethologica, 25*, 33–42. https://doi.org/10.1007/s10211-021-00379-6

Bates, D., Mächler, M., Bolker, B., & Walker, S. (2015). Fitting linear mixed-effects models using lme4. *Journal of Statistical Software, 67*(1), 1–48. https://doi.org/10.18637/jss.v067.i01

Beron, A., Keshavamurthy, R., Boutelle, C., & Wallace, R. M. (2026). Higher free-roaming dog density sustains rabies virus transmission in Haiti. *Scientific Reports, 16*. https://doi.org/10.1038/s41598-026-35359-y

Bhalla, S. J., Kemmers, R., Vasques, A., & Vanak, A. (2021). 'Stray appetites': a socio-ecological analysis of free-ranging dogs living alongside human communities in Bangalore, India. *Urban Ecosystems, 24*(6), 1245–1258. https://doi.org/10.1007/s11252-021-01097-4

Bhattacharjee, D., Sau, S., Das, J., & Bhadra, A. (2017). Free-ranging dogs prefer petting over food in repeated interactions with unfamiliar humans. *Journal of Experimental Biology, 220*(24), 4654–4660. https://doi.org/10.1242/jeb.166371

Bhattacharjee, D., Sau, S., & Bhadra, A. (2020). 'Bolder' together—Response to human social cues in groups of free-ranging dogs. *Behaviour, 157*(3–4), 363–384. https://doi.org/10.1163/1568539x-bja10005

Bhattacharjee, D., Sarkar, R., Sau, S., & Bhadra, A. (2021). Sociability of Indian free-ranging dogs (*Canis lupus familiaris*) varies with human movement in urban areas. *Journal of Comparative Psychology, 135*(1), 89. https://doi.org/10.1037/com0000241

Biswas, S., Ghosh, K., Touhid, S., Nandi, S., Bhattacharyya, A., Bhattacharyya, A., Das, M., Paul, R. G., & Bhadra, A. (2025). How Much Is Enough? An Empirical Test of the Resource Dispersion Hypothesis. *Ecology and Evolution, 15*, e72696-e72696. https://doi.org/10.1002/ece3.72696

Biswas, S., Nandi, S., Pal, T. S., Lahiri, A., Roy, A., Gope, H., Ghosh, K., Touhid, S., Ghosh, S., & Bhadra, A. (2026). Begging for morsels: social context and sex shape begging in free-ranging dogs. *Animal Behaviour*. https://doi.org/10.1016/j.anbehav.2026.123693

Boitani, L., & Ciucci, P. (1995). Comparative social ecology of feral dogs and wolves. *Ethology Ecology & Evolution, 7*(1), 49–72. https://doi.org/10.1080/08927014.1995.9522969

Bonanni, R., & Cafazzo, S. (2014). The social organisation of a population of free-ranging dogs in a suburban area of Rome: A reassessment of the effects of domestication on dogs' behaviour. In J. Kaminski & S. Marshall-Pescini (Eds.), *The social dog: Behavior and cognition* (pp. 65–104). Academic Press.

Bombieri, G., Del Mar Delgado, M., Russo, L., Garrote, P. J., López-Bao, J., Fedriani, J. M., & Penteriani, V. (2018). Patterns of wild carnivore attacks on humans in urban areas. *Scientific Reports, 8*. https://doi.org/10.1038/s41598-018-36034-7

Butler, J., Brown, W., & Du Toit, J. D. (2018). Anthropogenic food subsidy to a commensal carnivore: The value and supply of human faeces in the diet of free-ranging dogs. *Animals, 8*(5), 67. https://doi.org/10.3390/ani8050067

Büyük, E., Sözer, A. N., Yıldız, F., & Naderi, M. (2025). Crossing paths in fragmented urban forest ecosystems: a study on the pressure of stray dogs and humans on the activities of wild mammals. *Urban Ecosystems, 28*. https://doi.org/10.1007/s11252-025-01827-y

Chand, M., S., F. K., Yousuf, P., & Nautiyal, R. (2026). Coexistence or conflict? Understanding human–macaque encounters in Delhi's urban sanctuary. *Urban Ecosystems, 29*. https://doi.org/10.1007/s11252-025-01894-1

Chandran, E., & Azeez, P. (2016). Stray dog menace: Implications and management. *Economic and Political Weekly, 51*, 58–65.

Finzi, A., Rava, E., & D'Aniello, B. (2023). The begging strategy of Andean dogs: An exploratory study. *Animals, 13*(4), 704. https://doi.org/10.3390/ani13040704

Gaynor, K., Hojnowski, C., Carter, N., & Brashares, J. (2018). The influence of human disturbance on wildlife nocturnality. *Science, 360*, 1232–1235. https://doi.org/10.1126/science.aar7121

Gompper, M. E. (2014). The dog–human–wildlife interface: Assessing the scope of the problem. In M. E. Gompper (Ed.), *Free-ranging dogs and wildlife conservation* (1st ed., pp. 9–54). Oxford University Press.

Hartig, F. (2024). *DHARMa: Residual diagnostics for hierarchical (multi-level/mixed) regression models* (R package version 0.4.7) [Computer software]. https://CRAN.R-project.org/package=DHARMa

Hughes, J., & Macdonald, D. W. (2013). A review of the interactions between free-roaming domestic dogs and wildlife. *Biological Conservation, 157*, 341–351. https://doi.org/10.1016/j.biocon.2012.07.005

Kumar, N. (2026). The Indian street dog crisis and multispecies coexistence in tropical urban futures. *Nature Cities, 3*, 679-687. https://doi.org/10.1038/s44284-026-00476-2

Kumar, N., Jhala, Y., Qureshi, Q., Gosler, A., & Sergio, F. (2019). Human-attacks by an urban raptor are tied to human subsidies and religious practices. *Scientific Reports, 9*. https://doi.org/10.1038/s41598-019-38662-z

Lenth, R., & Piaskowski, J. (2026). *emmeans: Estimated marginal means, aka least-squares means* (R package version 2.0.2) [Computer software]. https://rvlenth.github.io/emmeans/

Lüdecke, D., Ben-Shachar, M., Patil, I., Waggoner, P., & Makowski, D. (2021). performance: An R package for assessment, comparison and testing of statistical models. *Journal of Open Source Software, 6*(60), 3139. https://doi.org/10.21105/joss.03139

Nandi, S., Lahiri, A., Pal, T. S., Roy, A., Bairagya, R., & Bhadra, A. (2026). Treats or affection? Understanding reward preferences in Indian free-ranging dogs. *Animal Cognition, 29*, 31. https://doi.org/10.1007/s10071-026-02046-4

Paul, M., & Bhadra, A. (2018). The great Indian joint families of free-ranging dogs. *PLoS ONE, 13*. https://doi.org/10.1371/journal.pone.0197328

Procko, M., Naidoo, R., LeMay, V., & Burton, A. C. (2023). Human presence and infrastructure impact wildlife nocturnality differently across an assemblage of mammalian species. *PLOS ONE, 18*. https://doi.org/10.1371/journal.pone.0286131

Ranjan, V., Badola, R., Hussain, S. A., & Dhakate, P. M. (2025). Assessing the implications of habitat transformations on human-large carnivore interactions outside protected areas. *Scientific Reports, 15*, 28784-28784. https://doi.org/10.1038/s41598-025-13808-4

R Core Team. (2023). *R: A language and environment for statistical computing*. R Foundation for Statistical Computing. https://www.R-project.org/

Ratsitorahina, M., Rasambainarivo, J. H., Raharimanana, S., et al. (2009). Dog ecology and demography in Antananarivo. *BMC Veterinary Research, 5*, 21. https://doi.org/10.1186/1746-6148-5-21

Reese, J. F. (2005). Dogs and dog control in developing countries. In D. J. Salem & A. N. Rowan (Eds.), *The state of the animals III: 2005* (pp. 55–64). Humane Society Press.

Schell, C. J., Stanton, L., Young, J., Angeloni, L., Lambert, J., Breck, S., & Murray, M. (2020). The evolutionary consequences of human–wildlife conflict in cities. *Evolutionary Applications, 14*, 178–197. https://doi.org/10.1111/eva.13131

Majumder, S. S., Bhadra, A., Ghosh, A., Mitra, S., Bhattacharjee, D., Chatterjee, J., Nandi, A. K., & Bhadra, A. (2013). To be or not to be social: foraging associations of free-ranging dogs in an urban ecosystem. *acta ethologica, 17*, 1-8. https://doi.org/10.1007/s10211-013-0158-0

Majumder, S. S., Paul, M., Sau, S., & Bhadra, A. (2016). Denning habits of free-ranging dogs reveal preference for human proximity. *Scientific Reports, 6*, 32014-32014. https://doi.org/10.1038/srep32014

Prasad, S. (2026). Rules Without Law: India's Supreme Court and the Governance of Street Dogs. *The Global Journal of Animal Law, 14*, 43-72. https://doi.org/10.71389/c9ms8344

Serpell, J. (Ed.). (1995). *The domestic dog: Its evolution, behaviour and interactions with people*. Cambridge University Press.

Silva, L. C., Friker, B., Warembourg, C., Kanankege, K. S. T., Wera, E., Berger-González, M., Alvarez, D. A., & Dürr, S. (2022). Habitat selection by free-roaming domestic dogs in rabies endemic countries

in rural and urban settings. *Scientific Reports, 12*, 20928-20928. https://doi.org/10.1038/s41598-022-25038-z

Sudarshan, M., Mahendra, B. J., Madhusudana, S., Narayana, D. A. H., Rahman, A., Rao, N., Meslin, F.-X., Lobo, D., Ravikumar, K., & Gangaboraiah. (2006). An epidemiological study of animal bites in India: Results of a WHO sponsored national multi-centric rabies survey. *The Journal of Communicable Diseases, 38*(1), 32–39.

The New Indian Express. (2025, August 25). Surge in stray dog attacks a major concern. https://www.newindianexpress.com/states/andhra-pradesh/2025/Aug/25/surge-in-stray-dog-attacks-a-major-concern

Venables, W. N., & Ripley, B. D. (2002). *Modern applied statistics with S* (4th ed.). Springer. https://www.stats.ox.ac.uk/pub/MASS4/

Wickham, H. (2016). *ggplot2: Elegant graphics for data analysis*. Springer-Verlag. https://ggplot2.tidyverse.org

**Supplementary information**

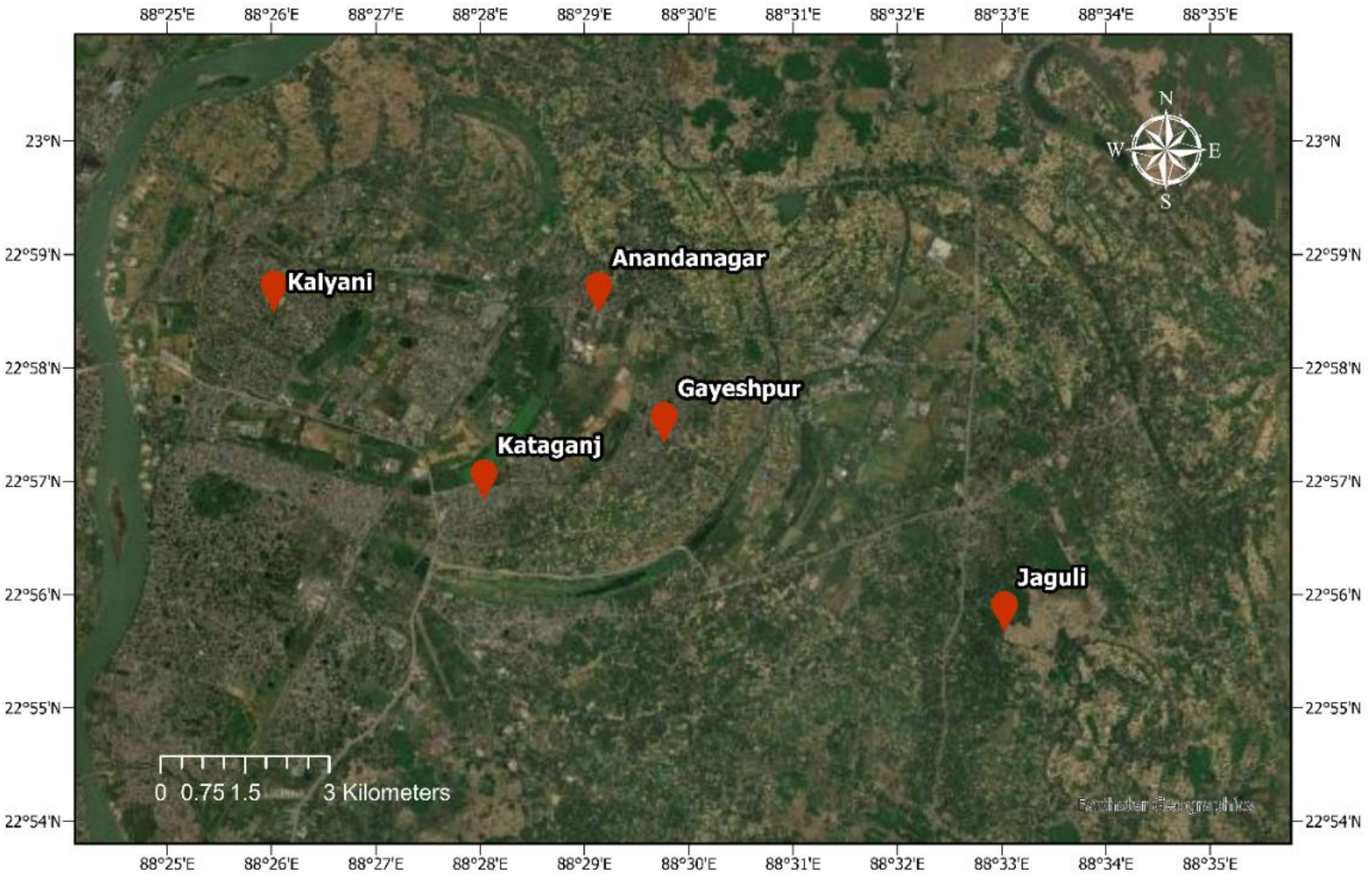

**Figure S1.** Map showing the locations of the study areas.

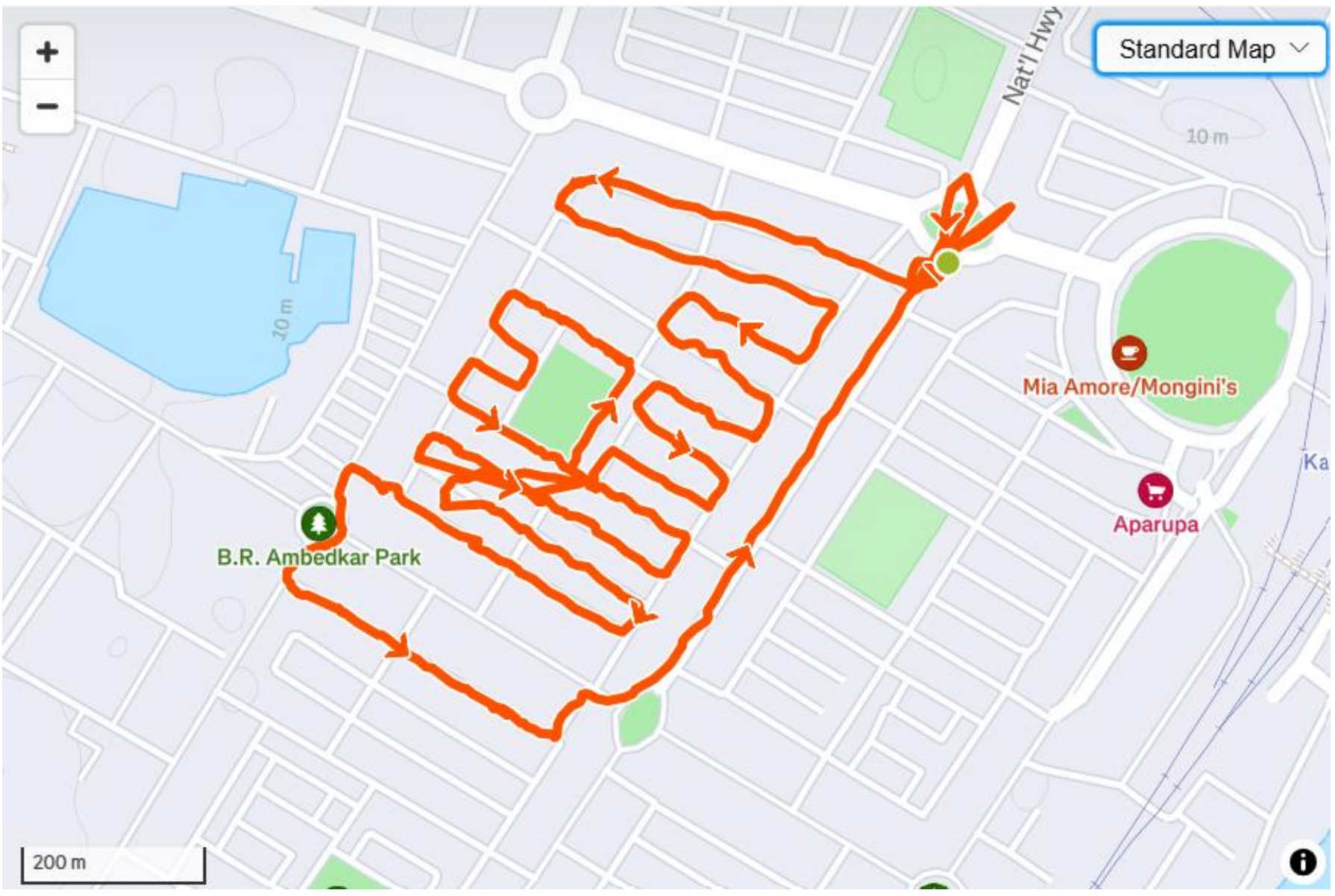

**Figure S2.** A snapshot of the location where the spot census was conducted. The orange lines depict the roads that were covered on foot during the spot census over two hours.

**Table S1.**

The number of surveys conducted in each area within a given location.

| **Location** | **Area** | **Number of surveys** |
| --- | --- | --- |
| Anandanagar | Area 1 | 21 |
| | Area 2 | 25 |
| | Area 3 | 23 |
| | Area 4 | 25 |
| Gayeshpur | Area 1 | 24 |
| | Area 2 | 25 |
| | Area 3 | 25 |
| | Area 4 | 25 |
| Jaguli | Area 1 | 25 |
| | Area 2 | 25 |
| | Area 3 | 24 |
| | Area 4 | 39 |
| Kalyani | Area 1 | 24 |
| | Area 2 | 20 |
| | Area 3 | 25 |
| | Area 4 | 25 |
| Kataganj | Area 1 | 25 |
| | Area 2 | 23 |
| | Area 3 | 27 |
| | Area 4 | 25 |

**Table S2.** Survey questionnaire.

| Question No. | Question | Options |
|---|---|---|
| 1 | Gender | Male / Female / Other |
| 2 | Age (in years) | Open-ended |
| 3 | Highest level of education | Never went to school / Up to Class 4 / Up to Class 10 / Up to Class 12 / Graduation / Post-Graduation and above |
| 4 | Do you have pets? | Yes / No |
| 5 | Do you feed stray dogs? | Yes / No |
| 6 | Do you pet stray dogs? | Yes / No |
| 7 | Do you take care of stray dogs when they fall sick? | Yes / No |
| 8 | Did you experience any negative interaction with stray dogs? | Yes / No |
| 9 | Were the dogs alone or in the presence of other dogs? | Alone / More than one dog |
| 10 | At what time of the day did you experience it? | 06:00-09:00 hours / 09:00-12:00 hours / 12:00-15:00 hours / 15:00-18:00 hours / 18:00-21:00 hours / 21:00-00:00 hours / 00:00-03:00 hours / 03:00-06:00 hours [More than one option can be chosen] |
| 11 | What did the dog do? | Open-ended |
| 12 | Do you think there is any diurnal difference in the aggressive behaviour of dogs towards humans? | Yes / No |
| 13 | If yes, at what time are they more aggressive? | 06:00-09:00 hours / 09:00-12:00 hours / 12:00-15:00 hours / 15:00-18:00 hours / 18:00-21:00 hours / 21:00-00:00 hours / 00:00-03:00 hours / 03:00-06:00 hours [More than one option can be chosen] |